\documentclass[journal,a4paper]{IEEEtran}

\usepackage{graphicx}
\usepackage{cite}
\usepackage{array}
\usepackage{amsmath,amssymb,amsfonts}
\usepackage{textpos}
\usepackage{subfigure}
\usepackage{enumerate}
\usepackage{lscape}
\usepackage{float}
\usepackage{verbatim}
\usepackage{pdfsync}
\usepackage{url}
\usepackage{color}
\usepackage{threeparttable}
\usepackage{multirow}
\usepackage{footnote}
\usepackage{textcomp}
\usepackage{xcolor}
\usepackage{bm}
\usepackage{makecell}
\usepackage{algorithm}
\usepackage{algpseudocode}
\usepackage{caption}

\begin{document}
\title{Deep Reinforcement Learning for Dynamic Spectrum Sensing and Aggregation in Multi-Channel Wireless Networks\
}

\author{Yunzeng~Li,
        Wensheng~Zhang,~\IEEEmembership{Member,~IEEE,}
        Cheng-Xiang~Wang,~\IEEEmembership{Fellow,~IEEE,}
        Jian~Sun,~\IEEEmembership{Member,~IEEE,}
        and Yu~Liu

\thanks{The authors acknowledge the support from the National Key {R\&D} Program of China under Grant 2018YFB1801101, the National Natural Science Foundation of China (NSFC) under Grant 61960206006, the Fundamental Research Funds of Shandong University under Grants 2017JC029 and 2017JC009, the China Scholarship Council (CSC) under Grant 201806225029, the High Level Innovation and Entrepreneurial Talent Introduction Program in Jiangsu, the Research Fund of National Mobile Communications Research Laboratory, Southeast University, under Grant 2020B01, the Fundamental Research Funds for the Central Universities under Grant 2242019R30001, Taishan Scholar Program of Shandong Province, the EU H2020 RISE TESTBED2 project under Grant 872172, and the Shandong Natural Science Foundation under Grant ZR2019BF040.}

\thanks{Y.~Li, W.~Zhang (corresponding author) and J.~Sun are with School of Information Science and Engineering, Shandong Provincial Key Lab of Wireless Communication Technologies, Shandong University, Qingdao, Shandong, 266237, China (e-mail: 17865197576@163.com; zhangwsh@sdu.edu.cn; sunjian@sdu.edu.cn).}

\thanks{C.-X.~Wang (corresponding author) is with National Mobile Communications Research Laboratory, School of Information Science and Engineering, Southeast University, Nanjing, Jiangsu, 210096, China, and Purple Mountain Laboratories, Nanjing 211111, China (e-mail: chxwang@seu.edu.cn).}

\thanks{Y.~Liu is with School of Microelectronics, Shandong University, Jinan, Shandong, 250101, China  (e-mail: yuliu@sdu.edu.cn).}
}
\markboth{IEEE Transactions on Cognitive Communications and Networking, ~Vol.~XX, No.~YY, MONTH~2020}%
\markboth{}%
\maketitle
\begin{abstract}
In this paper, the problem of dynamic spectrum sensing and aggregation is investigated in a wireless network containing $N$ correlated channels, where these channels are occupied or vacant following an unknown joint 2-state Markov model. {At each time slot, a single cognitive user with certain bandwidth requirement either stays idle or selects a segment comprising $C$ ($C<N$) contiguous channels to sense. Then, the vacant channels in the selected segment will be aggregated for satisfying the user requirement. The user receives a binary feedback signal indicating whether the transmission is successful or not (i.e., ACK signal) after each transmission, and makes next decision based on the sensing channel states.} Here, we aim to find a policy that can maximize the number of successful transmissions without interrupting the primary users {(PUs)}. The problem can be considered as a partially observable Markov decision process (POMDP) due to without full observation of system environment. We implement a Deep Q-Network (DQN) to address the challenge of unknown system dynamics and computational expenses. The performance of DQN, Q-Learning, and the Improvident Policy with known system dynamics is evaluated through simulations. The simulation results show that DQN can achieve near-optimal performance among different system scenarios only based on partial observations and ACK signals.
\end{abstract}

\begin{IEEEkeywords}
Dynamic spectrum aggregation, dynamic spectrum sensing, deep reinforcement learning, deep Q-network, POMDP.
\end{IEEEkeywords}

\section{Introduction}
\label{SEC_Introduction}
\begin{figure}[htb]
\centering
\includegraphics[width=0.9\columnwidth]{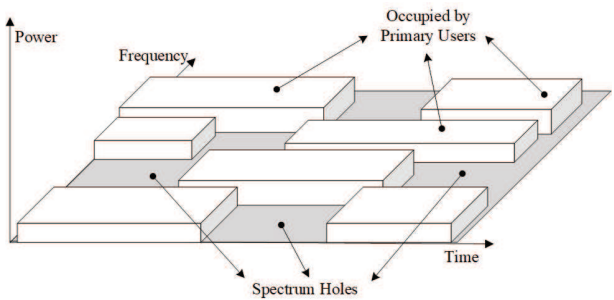}
\caption{The concept of spectrum holes.}
\label{SpectrumHoles}
\end{figure}

\begin{figure}[htb]
\centering
\includegraphics[width=0.9\columnwidth]{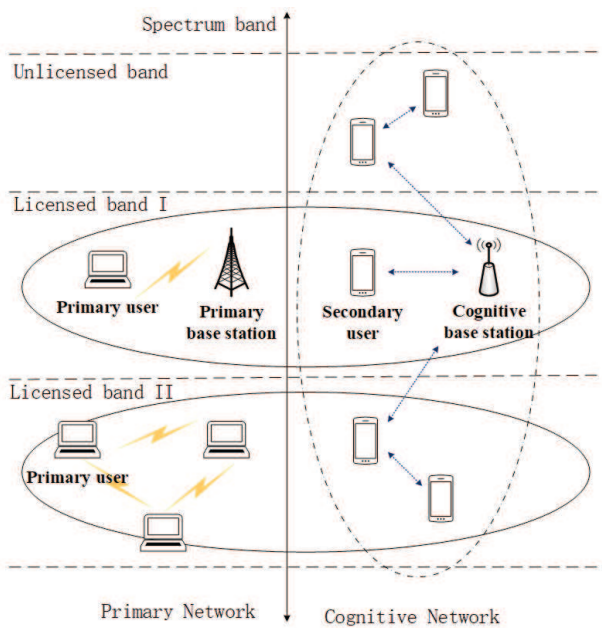}
\caption{{Structure of a cognitive wireless network.}}
\label{CRNET}
\end{figure}

In wireless networks{,} the spectrum is assigned to primary users (PUs) under a static and inflexible spectrum allocation policy, in which spectrum holes are not utilized in temporal or frequency domain as shown in Fig.~\ref{SpectrumHoles}. With the growing spectrum demand and limited spectrum resources, it is necessary to address the problem of spectrum underutilization and inefficiency. Cognitive radio \cite{paper1,paper2} has allowed secondary users (SUs) to sense and leverage the spectrum holes that are not occupied by PUs to improve spectrum utilization and alleviate spectrum scarcity. As shown in Fig.~\ref{CRNET}, there are two main parts in cognitive radio: primary network and cognitive network. PUs in the primary network are licensed to use spectrum bands, and SUs in the cognitive network have to access the spectrum holes in an opportunistic manner. 
The spectrum holes, however, are discrete and usually too insufficient to meet SUs' demand. {As a solution,} spectrum aggregation \cite{paper3,paper4} has attracted great concerns. Spectrum aggregation refers to the fact that a user can simultaneously access multiple discrete {spectrum} holes through Dis-contiguous Orthogonal Frequency Division Multiplexing (DOFDM) \cite{paper5} and aggregate them into a sufficiently wide band for successful transmission. {Although the aggregation capacity (i.e. the range of the aggregated bands) is fixed} due to the limitations of hardware \cite{paperCA,paperSA}, spectrum aggregation will play a critical and potential role in future cognitive radio networks \cite{Zhang_cognitive}.

\subsection{Existing Spectrum Occupancy Models}
\begin{figure*}[htbp]
\centering
\includegraphics[width=2\columnwidth]{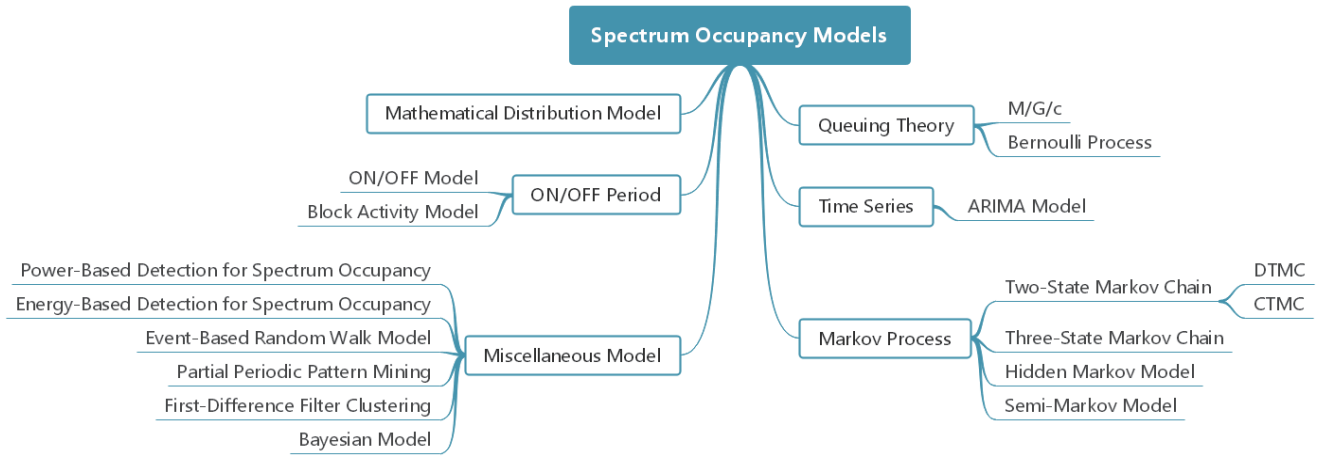}
\caption{Classification of existing spectrum occupancy models.}
\label{OccupancyModels}
\end{figure*}

Given that the spectrum occupancy activity of PUs leads to the dynamic and uncertainty of the spectrum environment, a reasonable spectrum occupancy model is necessary to {describe the channel state transition for utilization of spectrum holes}. The spectrum occupancy model can provide a reliable basis for the prediction of future spectrum occupancy status for SUs, thus the conflict between PUs and SUs will be
effectively reduced. Many spectrum occupancy models as shown in Fig.~\ref{OccupancyModels} have been discussed to simulate the behavior of PUs and describe the time-varying spectrum environment precisely.

\begin{figure}[htbp]
\centering
\includegraphics[width=1\columnwidth]{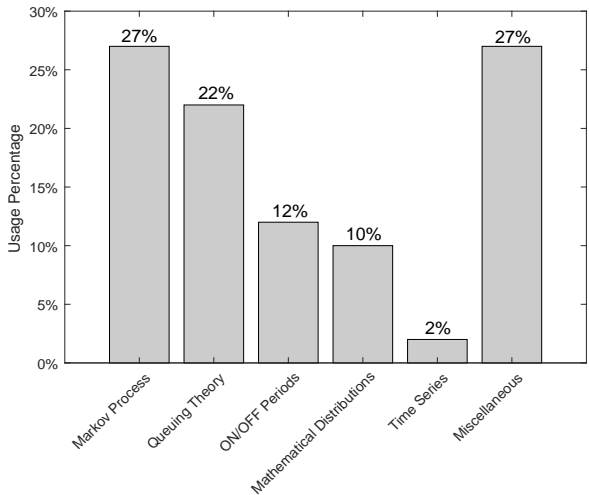}
\caption{Usage percentage of all spectrum occupancy models.}
\label{UsagePercentage}
\end{figure}

The usage percentage of these models is shown in Fig.~\ref{UsagePercentage} \cite{lopez2011overview,saleem2014primary}. These models can be divided into time-domain, frequency-domain and space-domain model from the perspective of spectrum measurement, or Markov process, Queuing theory, ON/OFF model, Time series, Mathematical distribution model and Miscellaneous model from the basis of modeling.
Different models focus on different characteristics of wireless spectrum environment, and there is no model completely applicable in various wireless scenarios. Since we focus on the state transition of the spectrum environment, the most widely used Markov model is adopted in this paper.

\subsection{Deep Reinforcement Learning and Related Works}
In recent years, Machine Learning (ML) has made great achievements, not only in computer vision and natural language processing \cite{nishani2017computer,lucas2018using,young2018recent,Wu2017}, but also in wireless communication \cite{bai2018predicting,wen2018deep,huang2018big,Chen2019}, incurring a collection of theoretical researches on optimization principle \cite{zhang2019standard,pennington2017geometry,louart2018random,nguyen2018loss,Wang2019}. As an important branch of ML, Reinforcement Learning (RL) is characterized by interacting with the changing and uncertain environment frequently to acquire knowledge, which provides an excellent performance in handling dynamic systems \cite{hu2018deep,liu2018deepnap,zhu2018deep}. Q-Learning implemented in this paper is one of popular RL methods. Instead of trying to model the dynamic characteristics of Markov decision process, Q-Learning directly estimates the Q-value of each action in each state. The Q-value estimates the expected accumulated discounted reward. The policy can then be executed by selecting the action with the highest Q-value for each state. Different from RL, Deep Reinforcement Learning (DRL) combines Deep Learning (DL) with RL, making it more capable of dealing with huge state space and complex computation.

Recently DRL has achieved significant breakthroughs in the dynamic spectrum allocation problems \cite{paper7,paper8,paper8trans,paper9,Chang2019Distributive,paper10,paper11,paper12,paper13}. The works in \cite{paper7},\cite{paper8},\cite{paper8trans},\cite{paper9} and \cite{Chang2019Distributive} studied the multichannel access problem under the assumption of Markov spectrum occupancy model. The authors of \cite{paper7} considered the highly correlation between channels thus the user can access the vacant channel by historical partial observations. A actor-critic DRL based framework was proposed in \cite{paper8,paper8trans} and its performance was further improved in \cite{paper7} especially in scenarios with a large number of channels. In \cite{paper9} all channels are independent so the user is supposed to have fully observation of the system via wideband spectrum sensing techniques. The independent channel model is also adopted in \cite{Chang2019Distributive}, but the authors of \cite{Chang2019Distributive} considered the presence of spectrum sensing errors, and the position of each user is specified in the proposed scenario. Moreover, multi-user scenarios are also studied in \cite{paper8trans} and \cite{Chang2019Distributive} through distributive learning. However, in the most of aforementioned works the user only selects one channel to access with the hope of avoiding collisions at each time slot. The authors of \cite{paper8trans} considered a scenario where the user can access more than one channel at a time, but it has nothing to do with the user's requirement. In other words, previous works mainly focused on allocating the single channel to the user without taking into account the user's demand of bandwidth.

Our simulation models are pretty different from the previous ones. What is new in our work is that we consider the user's requirement for broadband transmission (i.e. a successful transmission may be not affordable with a single channel) and provide the user sufficient bandwidth through applying the spectrum aggregation technology. In this paper, the correlation between channels is also taken into account, and the user will use this correlation to dispense with the perception of the whole frequency band. The user only needs to sense a segment in multiple channels, and the vacant channels in the selected segment will be aggregated for transmission. Meanwhile, the next segment to be sensed will be determined according to the sensing results. {The problem can be formulated as a partially observable Markov decision process (POMDP), where the user cannot accurately know the current state of the environment due to incomplete environmental observations.}

\subsection{Contributions}
We implement Deep Q-Network (DQN) \cite{paper14,fang2017learning} to approximate the action-value function which can give estimated Q-values of the user's available actions with {the partial observations of channel states} as input. We apply DQN into the dynamic spectrum sensing and aggregation problem in correlated channels to find a good policy to cope with the uncertain spectrum environment. The major contributions and novelties of our work can be summarized as follows:

\begin{itemize}
    \item {The bandwidth requirement of the user is considered in the correlated multichannel spectrum environment which is modeled as a Markov chain.} The user is given the spectrum aggregation capability to synthesize reliable frequency bands for successful transmission based on partial observations of the spectrum. We describe the problem as dynamic spectrum sensing and aggregation, and formulate the problem as a POMDP.

    \item DQN is adopted for the dynamic spectrum sensing and aggregation problem to deal with the uncertain spectrum environment. The action-value function is given by DQN through online learning to guide the user decision with no prior knowledge of system dynamics and low computational complexity.

    \item Q-Learning and the Improvident Policy based on full knowledge of the system is proposed to evaluate the performance of DQN. Simulations suggest that DQN can provide a near-optimal performance compared with Q-Learning and the Improvident Policy.
\end{itemize}

The rest of the paper is organized as follows. Section~\ref{SEC_Formulation} formulates the dynamic spectrum sensing and aggregation problem when channels are potentially correlated. Section~\ref{SEC_Framework} presents the Improvident Policy and the DQN framework to solve this problem, and Section~\ref{SEC_Simulation} shows through simulations that DQN can achieve near-optimal performance among different system scenarios. Finally, Section~\ref{SEC_Conclusion} concludes our work.

\section{Problem Formulation}
\label{SEC_Formulation}

The multichannel access problem has been studied in \cite{paper7,paper8,paper8trans,paper9,Chang2019Distributive} with single user where the state transition of channels is modeled as a joint Markov chain. The correlation between channels has been taken into account in \cite{paper7,paper8,paper8trans} while an independent channel model {has been} used in \cite{paper9,Chang2019Distributive}. The efficiency of spectrum aggregation and the spectrum assignment problem has been studied in \cite{paperCA} and \cite{paperSA}, respectively. The authors of \cite{paperCA,paperSA} considered the bandwidth demand of SUs and the fixed aggregation capability in multichannel network. Based on above works, we focus on the dynamic spectrum sensing and aggregation problem with one single user in {several} correlated channels. {Furthermore,} the user requirement for bandwidth and fixed aggregation capacity are taken into account in our work. In this section we formulate the problem in detail.

\subsection{System Model}
We consider a wireless network containing $N$ correlated channels whose states can be either vacant (0) or occupied (1). {The joint state transition of these channels} follows a $2^N$-states Markov model. Generally, the SU needs to sense the state of all channels and aggregate the vacant channels among them. However, due to the limited aggregation capability, only the vacant channels within the aggregation range can be utilized by the user, which makes the full-band sensing inefficient. Some existing works assume the dynamic radio environment as a simple independent-channel model, while in practice the external interference results in a high degree of correlation between these channels in wireless network \cite{paper7}. Based on the correlations between channels, we hope the single SU with certain bandwidth demand $d$ can only select a segment comprising $C$ channels to sense and aggregate the vacant channels for transmission, or just stay idle at the beginning of each time slot. The segment length is also the user's aggregation capability $C$ which is determined by the hardware limitations, thus all sensed vacant channels in the selected segment can be aggregated. If the number of the vacant channels in the selected segment is larger (smaller) than $d$, the transmission is successful (failed), which can be presented by ACK signal. ACK signal is the control character sent by the receiving station to the sending station through control channel, which indicates that the data has been received successfully. If the transmitter has obtained the ACK signal, it transmits the next block of data, otherwise it repeats the current block of data. Based on the sensing of the selected segment, the user decides what action to take in the next time slot. The goal is to achieve successful transmission as much as possible over time.


As the user can only sense the selected segment and has no full observations of the system, the problem can be formulated as a POMDP, where the user's environmental observation is incomplete at each time slot. Consequently, the user cannot even accurately know the current state of the system, and the prediction of the next state would be more difficult. Without the knowledge of the system dynamics, partial observations lead to larger state space and higher computational complexity. The user is supposed to deduce the current state from partial observations based on channel correlations, and infer the next state through learning from the Markov process.

\subsection{State Space}
\begin{figure}[htbp]

\centering
\includegraphics[width=0.9\columnwidth]{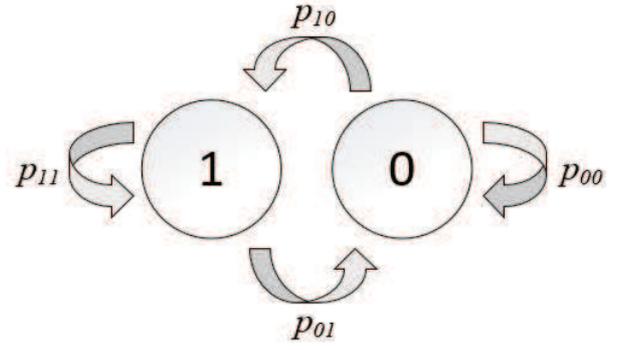}
\caption{State transition of each channel.}
\label{Makrov}
\end{figure}

\begin{figure*}[htbp]
\centering
\includegraphics[width=1.6\columnwidth]{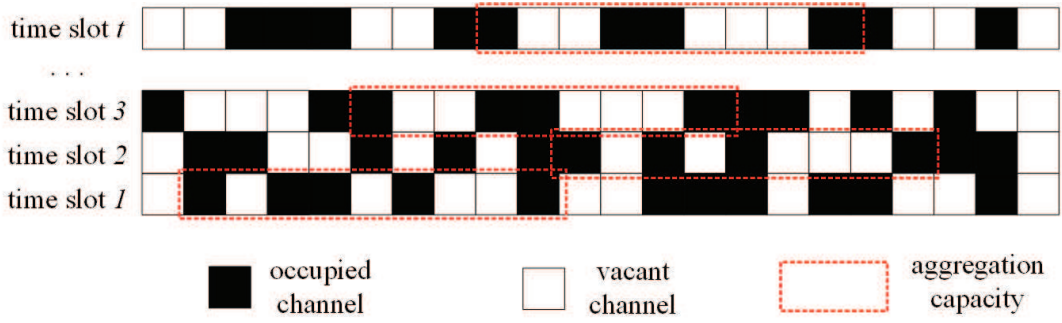}
\caption{The system spectrum environment.}
\label{environment}
\end{figure*}

Consider a wireless network with $N$ correlated channels divided from a shared bandwidth. Given that each channel has two possible states: occupied (1) or vacant (0), the whole system can be described as a $2^N$-states Markov model, and the state space is denoted as $S$

\begin{eqnarray}
     S=\lbrace\bm{s}=(s_1,...,s_N)|s_i\in\lbrace0,1\rbrace,i\in\lbrace1,...,N\rbrace\rbrace.
\end{eqnarray} Let $
\mathbf{P}=\left[\begin{array}{ll}{p_{00}} & {p_{10}} \\ {p_{01}} & {p_{11}}\end{array}\right]
$ be the transition matrix of the Markov chain and the state transition of each channel is shown in Fig.~\ref{Makrov}.
{The dynamic change of the spectrum environment with time in the whole system is shown in Fig.~\ref{environment}.}


\subsection{Action Space and User Observation}
A single user with certain bandwidth demand $d$ (i.e. it requires $d$ vacant channels for broadband transmission) is able to aggregate vacant channels in the range of aggregation capacity $C (C<N)$.
At the beginning of each time slot, the user either stays idle and transmits nothing or selects a length-$C$ segment of the whole channels to sense, and then there are $N-C+1$ segments for selection. The vacant channels in the selected segment will be aggregated for transmission. Let $A=\lbrace0,1,...,N-C+1\rbrace$ present the action space and the user will choose the $i^{th}$ segment at the beginning of time slot $t$ if $a_t=i (i\in{A},i\neq0)$ or transmit nothing and sense the state of the first segment acquiescently if $a_t=0$. The spectrum sensing errors are not taken into account. What the user observed, denoted as $\bm{o}_t\in\{(o_t^1,...,o_t^C)|o_t^i\in\{0,1\},i\in\{1,...,C\}\}$, is the basis to determine the next action $a_{t+1}$. Since the user only sense the state of selected segment, namely the whole system is partially observable to the user, the problem falls into a general POMDP. However, the user can use the correlation between channels to infer the current system state based on its decision and observation, and then deduce the next state and determine the next decision. If the number of vacant channels in the selected segment is larger than user demand $d$, the transmission is successful, otherwise failed.

\subsection{Reward Design}
After action $a_t$ is taken at time slot $t$, we assume that the user can receive a absolutely accurate binary feedback $f_t$ indicating whether its packet is successfully delivered (e.g. ACK signal). Assume $f_t=1$ if the transmission has succeed, otherwise $f_t=0$, and then we define reward function at time slot $t$ as

\begin{eqnarray}
\label{2}
             r_t(\bm{s}_t,a_t)=
             \begin{cases}
             0,& if \quad{a_t}=0  \\
             4f_t-2,& if\quad 1\le{a_t}\le{N-C+1}
             \end{cases}
\end{eqnarray}
where $s_t$ is system state at time slot $t$, which is not completely observable to the user but determines the binary feedback $f_t$ potentially.
Our objective is to find a policy $\pi$, which is a function mapping the observation $o_t$ to next action $a_{t+1}$ at each time slot, to maximize the excepted accumulated discounted reward

\begin{eqnarray}
V_{\pi}(\boldsymbol{o})=\mathbb{E}_{\pi}\left[\sum_{t=0}^{\infty} \gamma^{t} r_{t+1}\left(\boldsymbol{s}_{t+1}, \pi\left(\boldsymbol{o}_{t}\right)\right) | \boldsymbol{o}_{0}=\boldsymbol{o}\right]
\end{eqnarray}
where $\gamma \in{(0,1)}$ is a discount factor, $\pi{(o_t )}$ is the action under policy $\pi$ at time slot $t+1$ when current observation is $o_t$. The optimal policy ${\pi}^*$ can be presented as

\begin{eqnarray}
\begin{aligned} \pi^{*} &=\arg \max _{\pi} V_{\pi}(\boldsymbol{o}) \\
 &=\arg \max _{\pi} \mathbb{E}_{\pi}\left[\sum_{t=0}^{\infty} \gamma^{t} r_{t+1}\left(\boldsymbol{s}_{t+1}, \pi\left(\boldsymbol{o}_{t}\right)\right) | \boldsymbol{o}_{0}=\boldsymbol{o}\right]. \end{aligned}
\end{eqnarray}

\section{Improvident Policy and DRL Framework}
\label{SEC_Framework}
There are two approaches to cope with the dynamic spectrum sensing and aggregation problem in correlated channels formulated in Section~II: i) investigate the system transition matrix and make decisions based on the prior knowledge of the system dynamics, which is known as model-based planning; ii) approximate the function mapping observations to optimal action by interacting with the system directly. In this section, we propose the Improvident Policy with known full knowledge of the system dynamics as the first approach to obtain near-optimal performance. Q-Learning and DQN are adopted without any prior knowledge of the system dynamics as the second approach.

\subsection{Improvident Policy}

The Improvident Policy aims at maximizing immediate expected reward, which means that prior knowledge of the system dynamics is necessary. We assume that the system transition matrix has been known and the current state can be deduced precisely through partial observation under the Improvident Policy, and then the policy can be presented as

\begin{eqnarray}
\widehat{\pi}=\arg \max _{a} \sum_{s^{\prime} \in S} P\left(\boldsymbol{s}^{\prime} | \boldsymbol{s}\right) r\left(\boldsymbol{s}^{\prime}, a\right)
\end{eqnarray}
where $P\left(\boldsymbol{s}^{\prime} | \boldsymbol{s}\right)$ is the joint transition possibility from {current state $\bm{s}$ to next state $\boldsymbol{s}^{\prime}$}. Given next possible state $\boldsymbol{s}^{\prime}$ and action $a$, the reward $r\left(\boldsymbol{s}^{\prime}, a\right)$ is obtained by checking if the number of vacant channels in selected segment {satisfies} the user demand $d$, but not through ACK signal.

Note that the next state is only relevant to current state and has no business with what action has been taken, so the performance of the Improvident Policy with known system dynamics is near-optimal. The Improvident Policy is primarily designed to measure the performance of DQN, so we give it a lot of favorable assumptions. However, it is hard to obtain the system dynamics and infer the current state according to partial observation in practice.

\subsection{Q-Learning}

RL is an important branch of ML, which mainly includes four elements: agent, environment state, action and reward. The agent acts by observing the state of the environment and {obtains} rewards. According to the rewards, the agent gradually acquires the action strategy that adapts to the current environment. Therefore, RL is very suitable for solving the continuous decision-making problems in Markov process.

Though the dynamic spectrum sensing and aggregation problem turns into a POMDP due to partial observability of the whole system, we can convert the POMDP into a Markov decision process (MDP) by regarding {$\boldsymbol{x}$} as the system state where {$\boldsymbol{x}$} includes two parts: sensing action and corresponding observation of the sensed segment.

\begin{figure}[htbp]
\centering
\includegraphics[width=0.7\columnwidth]{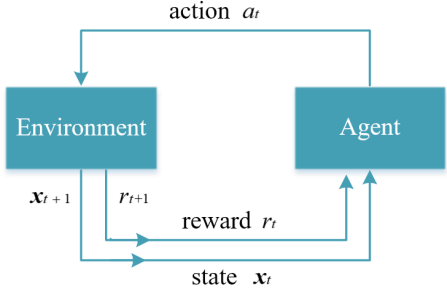}
\caption{Interaction between agent and environment in reinforcement learning.}
\label{interaction}
\end{figure}

In RL the agent interacts with the environment in discrete time, receives a reward $r$ corresponding to each state-action pair $({\boldsymbol{x}},a)$ as shown in Fig.~\ref{interaction}. Q-Learning as one of the most popular RL method aims at finding a sequence of actions to maximize the expected accumulated discounted reward through approximating an action-value function. The Q-value of the action $a$ under the state {$\boldsymbol{x}$}, given by the action-value function, denotes the expected revenue of the state-action pair $({\boldsymbol{x}}, a)$ {and the action with the largest Q-value will be chosen at each time slot.} We define $Q^{\pi}({\boldsymbol{x}}, a)$ as the action-value function when a sensing action {$a$} is taken in environment state {$\boldsymbol{x}$} under policy $\pi$. The Q-value of each state-action pair $({\boldsymbol{x}}, a)$, denoted as $q({\boldsymbol{x}}, a)$, updates through interacting with the system environment as follows

\begin{eqnarray}
\begin{aligned} &q\left(\boldsymbol{x}_{t}, a_{t+1}\right) \leftarrow  q\left(\boldsymbol{x}_{t+1}, a_{t+1}\right) \\ &+\alpha\left[r_{t+1}+\gamma \max _{a^{\prime} \in A} q\left(\boldsymbol{x}_{t+1}, a^{\prime}\right)-q\left(\boldsymbol{x}_{t}, a_{t+1}\right)\right] \end{aligned}
\end{eqnarray}
where $\alpha \in(0,1]$ is the learning rate.

The problem of Q-Learning is that the Q-value of each state-action pair is stored in a look-up
table. The large system state space leads to the scale of Q-value table increases enormously. As
a consequence, the Q-values of some state-action pairs cannot be sufficiently updated or even
seldom updated in limited iterations.

\subsection{Deep Q-Network}

\begin{figure}[htbp]
\centering
\includegraphics[width=0.5\textwidth]{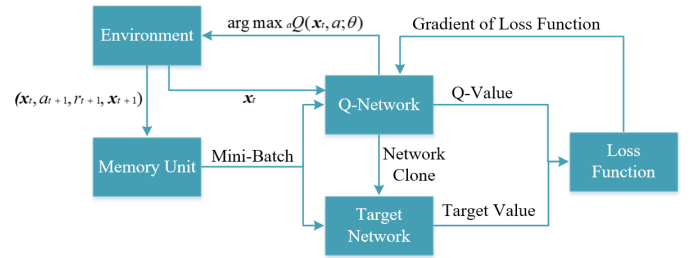}
\caption{An illustration of DQN.}
\label{DQN}
\end{figure}

{The performance of traditional RL methods is limited by the scale of the state space and action space of the problem. However, complex and realistic tasks are characterized by large state space and continuous action space, which are intractable with general RL methods. DRL combines DL with RL to enable the agent to deal with complex states and actions and DQN is one of the most popular DRL method. In DQN a deep neural network is adopted to replace the look-up table in Q-Learning to provide the Q-value of each state-action pair.} The neural network enables DQN to tackle the curse of dimensionality resulting from large system state space which is intractable with the Q-value table in Q-Learning. The main process of DQN for dynamic spectrum sensing and aggregation is detailed in Alg.~\ref{A_DSA}. The structure of DQN is presented in Fig.~\ref{DQN} and each component is specified as below.

\subsubsection{Input Layer}
The input of DQN is state $\boldsymbol{x}_{t}$ including the sensing action taken at time slot $t$ and corresponding observation, i.e., $\boldsymbol{x}_{t}=\left[\boldsymbol{a}_{t}, \boldsymbol{o}_{t}\right]$, where action vector $\boldsymbol{a}_{t}$ is the one-hot vector representation of action $a_t$ by setting the $({a_{t}+1})^{th}$ element to 1.

\subsubsection{Output Later}
The output of DQN is a vector of size $N-C+2$. The estimated Q-value if the user stays idle is presented in the first entry. The $({k+1})^{th}$ entry is the estimated Q-value for transmitting in the $k^{th}$ segment at time slot $t+1$, where $1\le{k}\le{N-C+1}$.

\subsubsection{Reward Definition}
The reward $r_t$ after taking action $a_t$ is obtained through ACK signal $f_t$, which has been defined in (\ref{2}).

\subsubsection{Q-Network}
The Q-network maps the current state to a string of action values, denoted as $Q\left(\boldsymbol{x}_{t}, a_{t+1} ; \theta\right) : \boldsymbol{x}_{t} \rightarrow\left\{q\left(\boldsymbol{x}_{t}, a ; \theta\right) | a \in A\right\}$, where $\theta$ is the parameters in the network. Given a state $\boldsymbol{x}$, the Q-values obtained from the Q-network present the estimates of the expected accumulated discounted rewards of all actions. After training process, the action with the largest Q-value will be taken in each time slot.

\subsubsection{Action Selection}
In the initial stage of training, {the Q-value of each state-action pair is not correct since the network has not converged.} If we take the action with the largest Q-value, most of the actions will not be implemented and the corresponding Q-values cannot be effectively updated. To get rid of the local optimum of DQN, $\epsilon$-greedy policy is adopted for action selection, where action $a_{t+1}=\arg \max _{a} Q\left(\boldsymbol{x}_{t}, a ; \theta\right)$ with possibility $1-{\epsilon}$ while a random action is selected with possibility $\epsilon$.

\begin{algorithm}[H]
\caption{\hspace{-0.5ex}: DQN algorithm for Dynamic Spectrum Sensing and Aggregation}\label{A_DSA}
\begin{algorithmic}

\State{$\bullet$ \textbf{Input:} memory size $M$, mini-batch size $B$, discount rate\\
\hspace{7ex}\quad $\gamma$, learning rate $\alpha$, $\epsilon$ in $\epsilon$-greedy policy, target\\
\hspace{7ex}\quad network update frequency $F$, and the number of\\
\hspace{7ex}\quad iterations $I_{max}$.}

\State{$\bullet$ \textbf{Do} \,\quad \,  Initialize the Q-network $Q\left(\boldsymbol{x}_{t}, a_{t+1} ; \theta\right)$ and its\\
 \hspace{7ex}\quad target network $\widehat{Q} \left(\boldsymbol{x}_{t}, a_{t+1} ; \hat{\theta}\right)$ with random \\
\hspace{7ex}\quad weights. Initialize the starting action $a_{0}$ and\\
\hspace{7ex}\quad execute it to get the initial state $\boldsymbol{x}_{0}$.\\
\hspace{7ex}\quad Initialize Train=\textbf{True}.}

\State{$\bullet$ \textbf{For} $t=1,2,...$ \textbf{do}\\
\quad\qquad  \textbf{If} Train \textbf{Then}\\
\quad\qquad \qquad Choose $a_{t}$ by $\epsilon$-greedy policy.\\
\quad\qquad \textbf{Else} \\
\quad\qquad \qquad  $a_{t}=\arg \max _{a} Q\left(\boldsymbol{x}_{t-1}, a ; \theta\right)$.  \\
\quad\qquad  \textbf{End If}\\
\quad\qquad Execute action $a_{t}$ and collect $r_{t}$ and $\boldsymbol{x}_{t}$.\\

\quad\qquad \textbf{If} Train \textbf{Then}\\
\quad\qquad \qquad Store ($\boldsymbol{x}_{t-1},a_{t},r_{t},\boldsymbol{x}_{t}$) in memory unit.\\
\quad\qquad\qquad \textbf{If} $t \geq M$ \textbf{Then}\\
\quad\qquad\qquad \qquad Remove the oldest experience tuple in\\
\quad\qquad\qquad \qquad memory unit.    \\
\quad\qquad\qquad \textbf{End If}\\
\quad\qquad \textbf{End If}\\
\quad\qquad  \textbf{If} Train \textbf{Then} \\
\quad\qquad \qquad Sample random mini-batch of experience tuples\\
\quad\qquad \qquad from memory unit. \\
\quad\qquad \qquad Compute the loss function $L(\theta)$ and update\\
\quad\qquad \qquad the weights $ \theta$.\\
\quad\qquad \qquad \textbf{If} $(t-M)\, mod\, F=0$ \textbf{Then}\\

\quad\qquad \qquad\qquad copy the weights  $\theta\rightarrow\hat{\theta}$. \\
\quad\qquad \qquad \textbf{If} $(t-M)>I_{max}$ \textbf{Then} Train$=$\textbf{False}\\
\quad \qquad  \textbf{End If}}
\State{$\bullet$ \textbf{End For}}\\
\hspace{-2ex}\textbf{Algorithm End} \hspace{38ex} $\Box$
\end{algorithmic}
\end{algorithm}

\subsubsection{Experience Replay}
For each time slot $t$, we refer to $\left(\boldsymbol{x}_{t}, a_{t+1}, r_{t+1}, \boldsymbol{x}_{t+1}\right)$ as an experience tuple stored in a memory unit and a mini-batch of experience tuples will be sampled in each iteration for training. As a supervised learning model, deep neural network requires the data to be independently and uniformly distributed, but the samples obtained through interaction with the environment are correlated. Experience Replay breaks the correlation by storage-sampling.

\subsubsection{Target Network}
We implement the target network, denoted as $\widehat{Q}\left(\boldsymbol{x}_{t}, a_{t+1} ; \hat{\theta}\right)$, to generate the target value. The target network has the same structure with the Q-network, whose parameters $ \hat{\theta}$ are copied from the Q-network at regular temporal intervals. In other words, The target network is updated at a lower frequency, which ensures that the target value received by Q-network in the training process is relatively stable.

\subsubsection{Loss Function}
 The loss function is defined as the mean square error of the target value and the Q-value, {i.e.,}

\begin{equation}
L(\theta)=\mathbb{E}\left[\left(y_{j}-Q\left(\boldsymbol{x}_{j}, a_{j+1} ; \theta\right)\right)^{2}\right]
\end{equation}

\noindent where $y_j$ is the target value combining the output of the target network and the reward

\begin{eqnarray}
y_{j}=r_{j+1}+\gamma \max _{a \in A} \widehat{Q}\left(\boldsymbol{x}_{j+1}, a ; \hat{\theta}\right).
\end{eqnarray}

As shown in {Alg.~\ref{A_DSA}}, different from general DL, there is no training data set or test data set for network training. In the proposed scheme, the user has to choose a channel segment for sensing and aggregation according to its bandwidth demand and aggregation capacity in a dynamic spectrum environment. Then the user receives a feedback (i.e. ACK signal) indicating whether the transmission is successful. We implement DRL to find a good selection strategy in the unknown dynamic environment. The user's selection of channel segments is treated as an interaction with the environment, and the reward is designed based on the ACK signal. The user simply interacts with the environment in a continuous way and learns from the rewards received in the process. When the learning process is finished, the user knows which actions should be issued in different environment states to obtain a greater reward. Therefore, it does not need to provide a special training set and test set, but only needs an interactive environment and accurate feedback. This is the reason why the training set is not required in the proposed scheme. During the learning process, the results of the user's interaction with the environment in the past period of time are stored in the memory unit as experience tuples, which will be taken out in batches for network training. The process in {RL} is called Experience Replay. Due to the experience tuples in the memory unit changes over time, it cannot be considered as a data set.

Note that the ACK signal can not only provide rewards during training process, but also serve as monitoring during DQN implementation. The performance degradation caused by the change of system environment will be reflected in the ACK signal, so that DQN can be reminded to enter the training process again. In other words, DQN can use ACK signal to adjust itself in time to better adapt to the dynamic spectrum environment. The whole process is shown in Alg.~\ref{RETRAIN_DSA}.


\begin{algorithm}[H]
\caption{\hspace{-0.5ex}: Re-training of DQN}\label{RETRAIN_DSA}
\begin{algorithmic}

\State{$\bullet$ \textbf{Do} Train the DQN and find a good policy as in Alg.~\ref{A_DSA}.}
\State{$\bullet$ \textbf{For} $t=1,2,...$ \textbf{do}\\
\quad\quad Perform actions according to the trained DQN and \\
\qquad receive rewards.\\
\qquad Calculate the accumulated reward.\\
\qquad  \textbf{If} the accumulated reward is less than a given threshold \\
\qquad \textbf{Then}\\
\quad \qquad Re-train DQN as in Alg.~\ref{A_DSA} to find a new policy.\\
\qquad  \textbf{End If}}
\State{$\bullet$ \textbf{End For}}\\

\hspace{-2ex}\textbf{Algorithm End} \hspace{38ex} $\Box$
\end{algorithmic}
\end{algorithm}

\section{{Simulation Results and Discussions}}
\label{SEC_Simulation}
In this section we compare DQN with other three policies: the Improvident Policy, Q-Learning, and the Random Policy. We assume that under the Improvident Policy the user can deduce the current states of the whole system channels precisely based on partial observation and the system transition pattern is also known, even though it is hard to achieve in practice. In the Random Policy, the user randomly selects one of available segments and aggregates the vacant channels for transmission at the beginning of each time slot.

\subsection{Details of DQN}
The neural network adopted in the DQN has four fully connected layers with each hidden layer containing 50 neurons and ReLU as activation function. ReLU is the most popular activation function, which is simple for computing and can avoid gradient explosion. Adam \cite{paper15} algorithm is applied to implement gradient descent during updating the parameters of the DQN, which is a commonly used optimizer with high convergence speed. The memory size is selected according to the number of possible system states and available user actions to ensure that the experience tuple of each state-action pair can be included. Mini-batch size is the number of samples fed into the model at each network training step. 32 is a commonly used value in {DL}, which can guarantee the fast convergence speed of the network and avoid the huge memory occupation. The $\epsilon$ in $\epsilon$-greedy policy is initially set as 0.9, and decreases to 0 in 10000 iterations. At the beginning of training, the network has not converged, so we encourage the user to explore the rewards brought by different actions. When the training is over, the network has converged, and the user can choose the action according to the given Q-value. Therefore, $\epsilon$ is supposed to gradually decrease from a value close to 1 during training. The discount rate (usually close to 1) is set to 0.9 so that the user will focus on rewards in the immediate future, because paying more attention to the long-term reward will make the training process slower and more difficult due to the uncertainty of long-term reward. Details of the DQN hyper-parameters are summarized in Table~\ref{Tab_Parameters}. The simulation is implemented in Tensorflow with GPU.

\begin{table}[htb]\normalsize  

\centering
\caption{Hyper-parameters of DQN.}
\label{Tab_Parameters}

  \begin{tabular}{| c | c |}
    \hline
    {Hyper-parameters} & {Value} \\\hline\hline
    {Memory size $M$}  & {$300000$}  \\\hline
    {Mini-batch size $B$} & {$32$}  \\\hline
    {Target network update frequency} & {200} \\ \hline
    {$\epsilon$ in $\epsilon$-greedy policy} & {${0.9}\rightarrow{0}$} \\\hline
    {Discount rate $\gamma$} & {$0.9$} \\\hline
    {Activation function} & {ReLU} \\\hline
    {Optimizer} & {Adam}\\\hline
    \end{tabular}
\end{table}

\subsection{Performance Evaluation}
We consider the system containing $N=24$ channels which are highly correlated, where several independent channels follow the same 2-state Markov chain with transition matrix $\mathbf{P}$, while the state of any other channel is the same (the correlation coefficient $\rho=1$) or opposite ($\rho=-1$) to that of an independent channel. Note that which channels are interrelated is randomly determined.

As discussed in Section~\ref{SEC_Formulation}, the user can stay idle $(a=0)$ or select one of available segments $(a=i, i>0)$ to sense, and then there are four possible situations after an action is taken: $i)$ $a=0$ and none of the segments in the system environment could satisfy the user's transmission demand, i.e., a successful transmission is impossible in current system state; $ii)$ $a=0$ but there exists available segments in which the number of vacant channels could afford a successful transmission; $iii)$ $a=i,i>0$ and the transmission succeeds; $iv)$ $a=i,i>0$ but the transmission fails, which means that the PUs are disrupted in a way. However, the user is not able to tell the first two situations in practice due to partial observation of the system, but the difference between them is taken into account in simulations to evaluate the performance. We assume the user has made the right decision if it achieves a successful transmission or stays idle when the system environment is terrible and calculate the decision accuracy in 10000 time slots.

\subsubsection{Decision accuracy}

\begin{figure}[htbp]
\centering
\includegraphics[width=0.9\columnwidth]{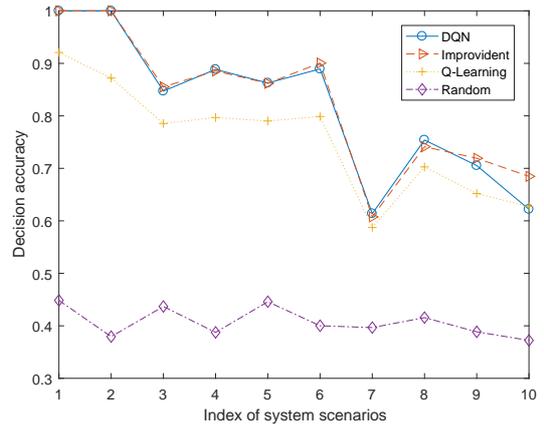}
\caption{Decision accuracy of four policies in 10 different scenarios when $C=8$ {and} $d=4$.}
\label{DA84}
\end{figure}

Fig.~\ref{DA84} shows the decision accuracy of four policies: DQN, Improvident Policy, Q-Learning, and Random Policy. We assume there are 4, 5, or 6 independent channels in the system and change the transition matrix  $\mathbf{P}$ to get 10 different dynamic system scenarios with the correlation coefficient $\rho=-1$ in the first six scenarios and $\rho=1$ in the last four scenarios. Specifically, there are 4 independent channels in scenario 1, 2, 7, and 8, 5 independent channels in scenario 3, 4, and 9, and 6 independent channels in scenario 5, 6, and 10. Moreover, four different transfer matrices are used in these ten scenarios. As shown in Fig.~\ref{DA84}, the Improvident Policy, which is assumed having full knowledge of the system dynamics, as well as DQN, performs better than Q-Learning. In spite of the scarcity of the system prior knowledge, DQN achieves a performance very close to the Improvident Policy in most scenarios, which indicates the strong learning ability of DQN in dynamic environment. Q-Learning works worse than DQN due to incapable of dealing with large state space.

Additionally, the performance curves of the three policies above has a larger fluctuation among all scenarios than the Random Policy, which indicates that the state transitions in different scenarios have different randomness, making the achievable performance limited even if the system dynamics is fully known. If the randomness of the state transition of each channel {were} very small (i.e. the probability of state transition is {too} large or {too} small), it would be easy for the user to predict the next state of the system environment, so as to achieve a high decision accuracy such as $100$$\%$ in scenario~1. If the randomness of the state transition of each channel {were} very large, the achievable optimal decision accuracy would be limited under any policies. Therefore in scenario~7 the decision accuracy of DQN, Improvident Policy, and Q-Learning is about $60$$\%$. Different scenarios are independent {with} each other in our simulations. In different scenarios, there are different number of states, different probability of state transition and different sets of relevant channels. So even under the same policy, the achievable optimal performance is diverse in different scenarios due to the limitation of environmental conditions. Therefore, the performance curve fluctuates with the scenario. However, as shown in the figure, the proposed DQN has close performances with the near-optimal Improvident Policy in various scenarios, indicating that {the proposed algorithm} can achieve near-optimal performance regardless of the spectrum environment.

\begin{figure}[htbp]
\centering
\includegraphics[width=0.9\columnwidth]{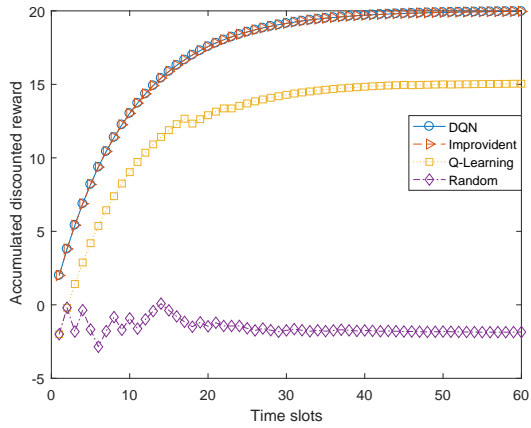}
\caption{Accumulated discounted reward of four policies in scenario~1 when $C=8$ {and} $d=4$.}
\label{DiscountedReward}
\end{figure}

Fig.~\ref{DiscountedReward} shows the accumulated discounted reward over time of four policies in scenario~1. It can be seen that except for the Random Policy, the accumulated discounted rewards of other policies increase gradually with time and finally stabilize. The curves of DQN and Improvident Policy are coincident and stable at the highest value, indicating that DQN achieves extremely close performance to Improvident Policy even in the absence of system dynamics information.

\begin{figure}[htbp]
\centering
\includegraphics[width=0.9\columnwidth]{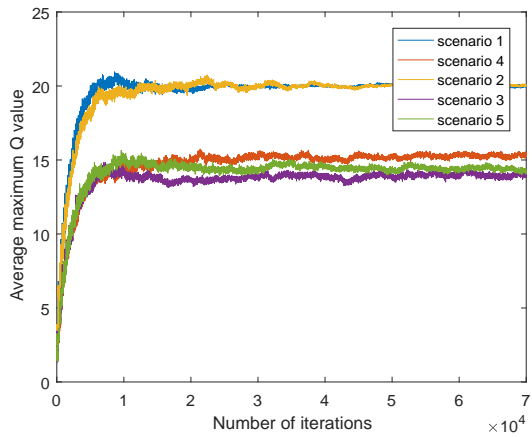}
\caption{Average maximum Q-values of DQN in training process when $C=8$ {and} $d=4$.}
\label{MaxQ}
\end{figure}

In addition, we give the change of maximum Q-value over time to show the learning process of DQN. The maximum Q-value for all actions in a given state represents the estimate of the maximum expected cumulative discount reward. We calculated the average maximum Q-value in each iteration from ten different initial states of the first five scenarios. As shown in the Fig.~\ref{MaxQ}, the average maximum Q-value in all scenarios increases and remains stable, indicating that DQN gradually learns a good policy and maintains it.

Previous works \cite{paper7,paper8,paper8trans,paper9} tended to adopt obtained reward for performance evaluation. In our work, we defined different performance evaluation parameters for four different situations as discussed above. It is necessary to distinguish situation $i)$ and $ii)$ accurately in performance evaluation because staying idle when the spectrum environment is in good condition is not the correct decision, but still better than a failed transmission, because transmission failure will lead to the interference to the PUs. However, these two situations cannot be differentiated through reward because staying idle means the obtained reward is 0. So we treat situation $i)$ and $iii)$ as the right decision, situation $ii)$ as the conservative decision, situation $iv)$ as the wrong decision, and assign different weights to them in the definition of the modified decision accuracy. We define the modified decision accuracy as


\begin{eqnarray}
\begin{aligned} &\text { modified decision accuracy }\\&=\frac{ \text { \#success }+\beta \times  \text { \#conservative }}{ \text { \#timeslots }}
\end{aligned}
\end{eqnarray}

\noindent where $ \text { \#success }$ is the number of times the user takes the correct decision, $ \text { \#conservative }$ is the number of times the user stays idle while the system environment is in good condition, $\beta$ is the measurement weight of such conservative choices, which is set as 0.5 in this paper, and $ \text { \#timeslots }$ is the total number of time slots.

\begin{figure}[htbp]
\centering
\includegraphics[width=0.9\columnwidth]{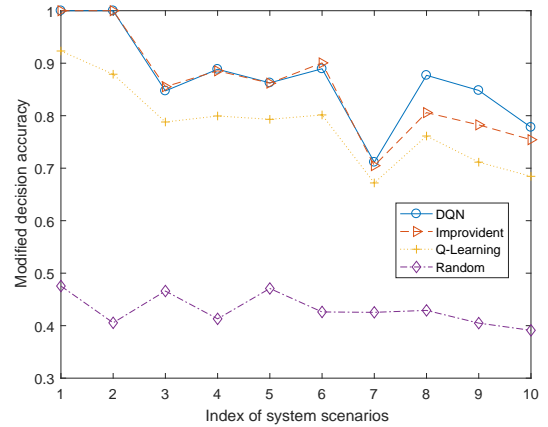}
\caption{Modified decision accuracy of four policies in 10 different scenarios when $C=8$ {and} $d=4$.}
\label{MDA84}
\end{figure}

In the definition of modified decision accuracy, the weights of the correct decision and the failed transmission can be viewed as 1 and 0, respectively. So the weight of $ \text { \#conservative }$ is supposed to be between 0 and 1, which depend on how much we approve of this conservative action. We regard the conservative action as half-correct decision so the weight of $ \text { \#conservative }$ is set to the middle value. The result is shown in Fig.~\ref{MDA84}.

\begin{figure}[htbp]
\centering
\includegraphics[width=0.9\columnwidth]{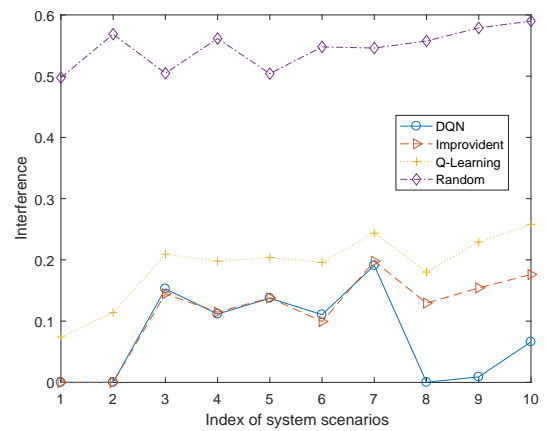}
\caption{Interference of four policies in 10 different scenarios when $C=8$ {and} $d=4$.}
\label{I84}
\end{figure}

Compared with Fig.~\ref{DA84}, the modified decision accuracy of DQN is higher than the decision accuracy in the last three scenarios, but in the other policies the change is little, which means that at some time slots the user of DQN stays idle and avoids transmission failure, consequently reduces the interference to PUs.

We can compare the interference directly of four policies resulting from transmission failure, and the interference is defined as

{
\begin{eqnarray}
\text{interference}=\frac{  \text { \#failure }}{  \text { \#timeslots }}
\end{eqnarray}}
where $ \text{ \#failure }$ is the number of failed transmission.

Fig.~\ref{I84} shows that compared with the Improvident Policy, the interference of DQN is similar or even lower, that is why the modified decision accuracy of DQN is better than that of the Improvident Policy in the last three scenarios.

The proposed DQN can be theoretically extended for cases with any user demand and aggregation capacity. However, the aggregation capacity and the user demand jointly determine the problem complexity of the dynamic spectrum sensing and aggregation. If the aggregation capability {were} strong enough and the user requirement {were} small, it would be easy to achieve successful transmissions under any policy. If the aggregation capability {were} weak and the user requirement {were} large, a successful transmission tends to be impossible. So we assume the aggregation capacity as 1/3 of the full-band and the user demand as 1/2 of the aggregation capacity to make the problem more reasonable and practical. We also compare the performance of four policies with different {values of} adjusted user demand and aggregation capacity to indicate that the proposed algorithm can perform well regardless of the value of user demand and aggregation capacity.

\begin{figure}[htbp]
\centering
\includegraphics[width=0.9\columnwidth]{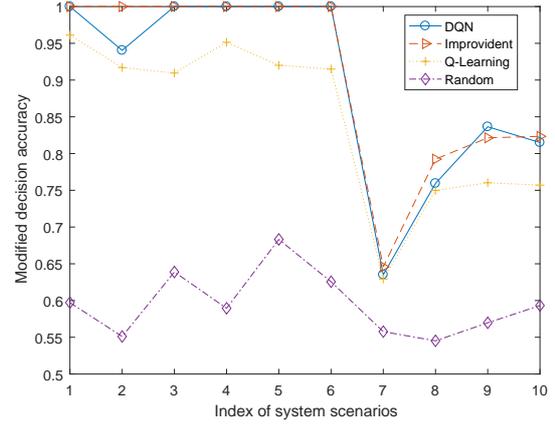}
\caption{Modified decision accuracy of four policies in 10 different scenarios when $C=8$ {and} $d=3$.}
\label{MA83}
\end{figure}

\begin{figure}[htbp]
\centering
\includegraphics[width=0.9\columnwidth]{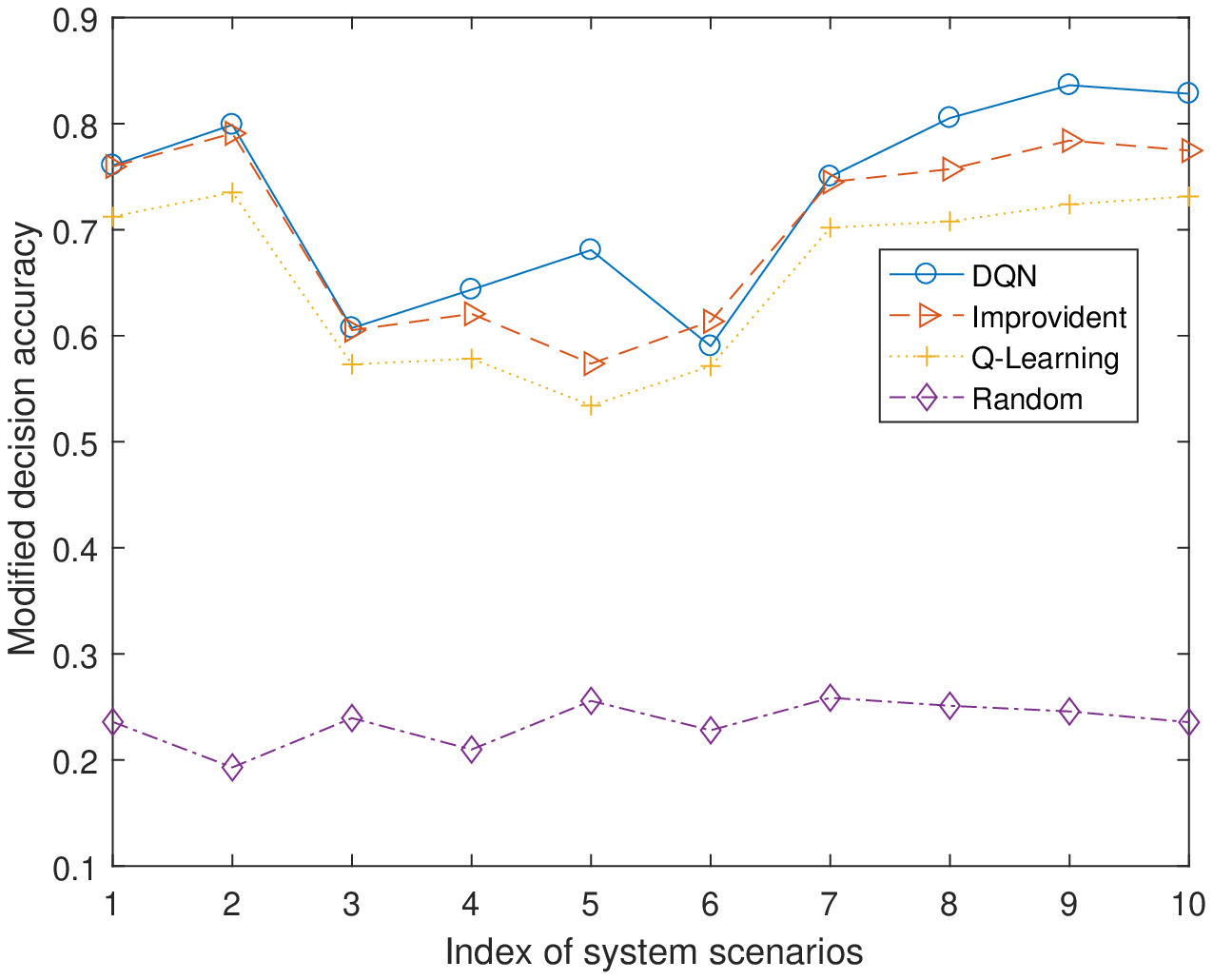}
\caption{Modified decision accuracy of four policies in 10 different scenarios when $C=8$ {and} $d=5$.}
\label{MA85}
\end{figure}

\begin{figure}[htbp]
\centering
\includegraphics[width=0.9\columnwidth]{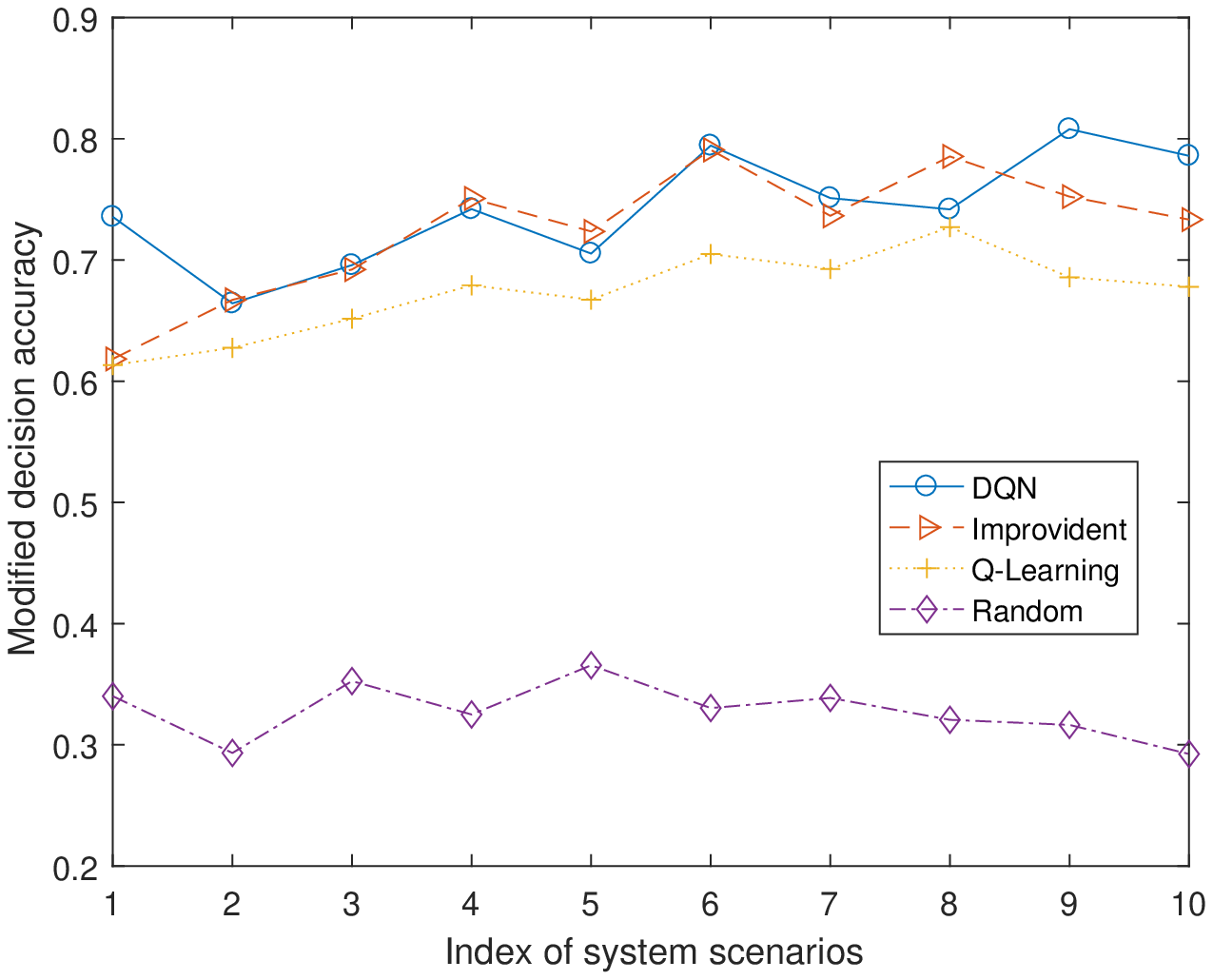}
\caption{Modified decision accuracy of four policies in 10 different scenarios when $C=7$ {and} $d=4$.}
\label{MA74}
\end{figure}

\begin{figure}[htbp]
\centering
\includegraphics[height=0.68\columnwidth,width=0.9\columnwidth]{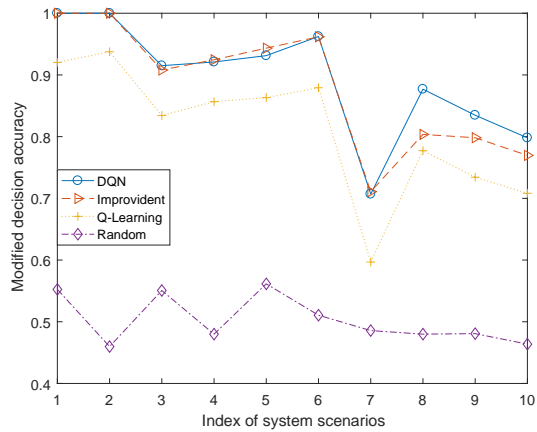}
\caption{Modified decision accuracy of four policies in 10 different scenarios when $C=9$ {and} $d=4$.}
\label{MA94}
\end{figure}

The robustness of DQN with different aggregation capacity $C$ and user demand $d$ is verified and the results are shown in Fig.~\ref{MA83}--\ref{MA94}. We found that in most scenarios DQN performs closely to the Improvident Policy even without any knowledge of the system dynamics. In some special scenarios DQN performs best, which mainly because DQN tends to stay idle to avoid transmission failure at some intractable time slots. In a very few scenarios, the performance of DQN is significantly worse than that of the Improvident Policy, but it is still better than that of Q-Learning.

\subsubsection{Computational complexity}

From the perspective of temporal complexity, DQN and Q-Learning have an obvious advantage because the mapping from state to action has been learned through training process, while the Improvident Policy needs to apply the knowledge of the system dynamics in each time slot. The Improvident Policy has to compute the transition possibilities from the current state to the next possible states, as well as the reward for each action in each of the next possible states. So the temporal complexity of the Improvident Policy can be presented as $O(2^i+2^i \times (N-C+2))$, where $i$ is the number of independent channels, $2^i$ denotes the number of the next possible states, and $N-C+2$ is the number of actions. Both DQN and Q-Learning have temporal complexity of $O(1)$ since there is no recurrent computation.

\begin{figure}[htbp]
\centering
\includegraphics[height=0.68\columnwidth,width=0.9\columnwidth]{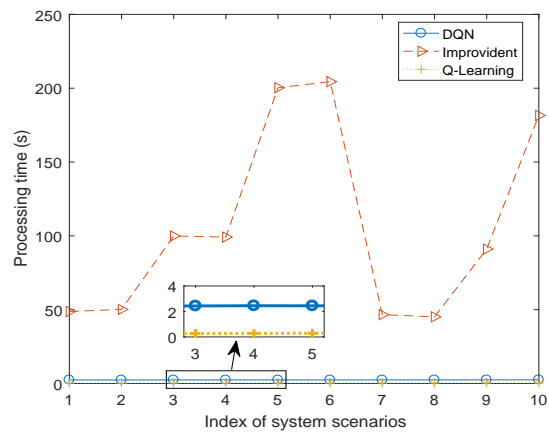}
\caption{Average processing time of three policies in 10 different scenarios.}
\label{time}
\end{figure}

In Fig.~\ref{time} we show the average processing time of three policies for 10000 time slots in our device to verify the aforementioned discussions. The Improvident Policy has significantly lower time efficiency than the other two strategies. What is worse is that the curve of the Improvident Policy varies greatly in different scenarios, because each additional independent channel doubles the number of the next possible state, and thus the processing time. Q-Learning definitely has the least processing time in that it uses table lookup instead of computation. DQN provides a performance very close to Q-Learning in temporal complexity. Although DQN and Q-Learning require extra training time compared with the Improvident Policy, they have greater advantages in the long run. By the way, Q-Learning with the lowest temporal complexity also has the highest spatial complexity due to the huge size of look-up table. Another advantage of DQN and Q-Learning over the Improvident Policy is that their curves barely fluctuate in different scenarios, which reflects the robustness of their performance in different system environment.

In terms of spatial complexity, since DQN has its fixed network structure, no matter how large the environment state space and action space are, the storage space of the algorithm cannot be affected. So the spatial complexity of DQN is $O(1)$. The Improvident Policy needs to traverse all possible next states to obtain the expected reward of each action, so its spatial complexity is proportional to the number of states, which can be formulated as $O(2^i)$. Although Q-Learning has the highest temporal efficiency than the other policies, it counts on a huge spatial complexity of $O(2^i \times (N-C+2))$ because Q-Learning has to store the Q-value corresponding to each action in each state. As discussed above, only the spatial complexity of DQN does not increase with the size of the problem. While the spatial complexity of Improvident Policy is affected by the state space, the spatial complexity of Q-Learning is affected by both the state space and the action space.

\section{Conclusions}
\label{SEC_Conclusion}
In this paper, we have considered the correlation between channels in {wireless networks} and modeled the dynamic spectrum environment as a joint Markov chain. We {have assumed that} the SU with certain bandwidth demand has the fixed aggregation capacity to access multiple vacant channels simultaneously for successful transmission. At each time slot, the SU either stays idle or selects a segment of the spectrum to sense. The segment length is determined by the aggregation capacity so that the SU can aggregate the vacant channels in the selected segment. The next user decision will be made based on the sensed state of the selected segment. We {have formulated} the dynamic spectrum sensing and aggregation problem as a POMDP and proposed a DQN framework to address it. We have compared the performance of three different policies: DQN, Q-Learning, and the Improvident Policy with known the system dynamics. Simulations have shown that the DQN can achieve a near-optimal decision accuracy in most system scenarios even without the prior knowledge of the system dynamics. The performance is also robust among {different aggregation capacities and different user demands for bandwidth}. Moreover, DQN has the lowest computational complexity {in both} time and space, and the temporal and spatial complexity of DQN is not affected by the expansion of the state space or action space of the problem.


\begin{thebibliography}{10}
\providecommand{\url}[1]{#1}
\csname url@samestyle\endcsname
\providecommand{\newblock}{\relax}
\providecommand{\bibinfo}[2]{#2}
\providecommand{\BIBentrySTDinterwordspacing}{\spaceskip=0pt\relax}
\providecommand{\BIBentryALTinterwordstretchfactor}{4}
\providecommand{\BIBentryALTinterwordspacing}{\spaceskip=\fontdimen2\font plus
\BIBentryALTinterwordstretchfactor\fontdimen3\font minus
  \fontdimen4\font\relax}
\providecommand{\BIBforeignlanguage}[2]{{%
\expandafter\ifx\csname l@#1\endcsname\relax
\typeout{** WARNING: IEEEtran.bst: No hyphenation pattern has been}%
\typeout{** loaded for the language `#1'. Using the pattern for}%
\typeout{** the default language instead.}%
\else
\language=\csname l@#1\endcsname
\fi
#2}}
\providecommand{\BIBdecl}{\relax}
\BIBdecl

\bibitem{paper1}
J.~Mitola, and G.~Q. Maguire,
  ``\BIBforeignlanguage{English}{Cognitive radio: making software radios more
  personal},'' \emph{\BIBforeignlanguage{English}{IEEE Pers. Commun.}}, vol.~6,
  no.~4, pp. 13--18, Aug. 1999.

\bibitem{paper2}
Y.-C. Liang, K.-C. Chen, G.~Y. Li, and P.~Mahonen,
  ``\BIBforeignlanguage{English}{Cognitive radio networking and communications:
  {A}n overview},'' \emph{\BIBforeignlanguage{English}{IEEE Trans. Veh.
  Technol.}}, vol.~60, no.~7, pp. 3386--3407, {Sept.} 2011.

\bibitem{paper3}
A.~Shukla, B.~Willamson, J.~Burns, E.~Burbidge, A.~Taylor, and D.~Robinson, ``A
  study for the provision of aggregation of frequency to provide wider
  bandwidth services,'' \emph{A report}, 2006.

\bibitem{paper4}
W.~Wang, Z.~Zhang, and A.~Huang, ``Spectrum aggregation: Overview and
  challenges{,}'' \emph{Network Protocols \& Algorithms}, vol.~2, no.~1, pp.
  184--196, May 2010.

\bibitem{paper5}
J.~D. Poston and W.~D. Horne, ``Discontiguous {OFDM} considerations for dynamic
  spectrum access in idle {TV} channels,'' in \emph{Proc. IEEE Int. Symp. New
  Front. Dyn. Spectr. Access Netw. (DySPAN)}, {Baltimore, USA,} Nov. 2005, pp. 607--610.

\bibitem{paperCA}
B.~{Gao}, Y.~{Yang}, and J.~{Park}, ``Channel aggregation in cognitive radio
  networks with practical considerations,'' in \emph{Proc. IEEE Int. Conf.
  Commun. (ICC)}, {Kyoto, Japan,} Jun. 2011, pp. 1--5.

\bibitem{paperSA}
F.~{Huang}, W.~{Wang}, H.~{Luo}, G.~{Yu}, and Z.~{Zhang}, ``Prediction-based
  spectrum aggregation with hardware limitation in cognitive radio networks,''
  in \emph{Proc. IEEE Veh. Technol. Conf. (VTC)}, {Taipei, China,} May 2010, pp. 1--5.

\bibitem{Zhang_cognitive}
W.~Zhang, C.-X. Wang, X.~Ge, and Y.~Chen, ``Enhanced 5{G} cognitive radio
  networks based on spectrum sharing and spectrum aggregation,'' \emph{IEEE
  Trans. Commun.}, vol.~66, no.~12, pp. 6304--6316, Dec. 2018.

\bibitem{lopez2011overview}
M.~L{\'o}pez-Ben{\'\i}tez and F.~Casadevall, ``An overview of spectrum
  occupancy models for cognitive radio networks,'' in \emph{Proc. International
  Conference on Research in Networking}.\hskip 1em plus 0.5em minus 0.4em\relax
  Springer, {Berlin, Germany,} 2011, pp. 32--41.

\bibitem{saleem2014primary}
Y.~Saleem and M.~H. Rehmani, ``Primary radio user activity models for cognitive
  radio networks: A survey,'' \emph{J. Netw. Comput. Appl.}, vol.~43, pp.
  1--16, {Aug.} 2014.



\bibitem{nishani2017computer}
E.~Nishani and B.~{\c{C}}i{\c{c}}o, ``Computer vision approaches based on deep
  learning and neural networks: Deep neural networks for video analysis of
  human pose estimation,'' in \emph{Proc. IEEE Mediterranean Conf. Embedded
  Comput. (MECO)}, {Bar, Montenegro,} Jun. 2017, pp. 1--4.


\bibitem{lucas2018using}
A.~Lucas, M.~Iliadis, R.~Molina, and A.~K. Katsaggelos, ``Using deep neural
  networks for inverse problems in imaging: Beyond analytical methods,''
  \emph{IEEE Signal Process. Mag.}, vol.~35, no.~1, pp. 20--36, Jan. 2018.

\bibitem{young2018recent}
T.~Young, D.~Hazarika, S.~Poria, and E.~Cambria, ``Recent trends in deep
  learning based natural language processing,'' \emph{IEEE Comput. Intell.
  Mag.}, vol.~13, no.~3, pp. 55--75, Aug. 2018.

\bibitem{Wu2017}
Y. Wu, F. Hu, G. Min, and A. Zomaya, \emph{Big data and computational intelligence in networking}, Florida: Taylor {\&} Francis, 2017.

\bibitem{bai2018predicting}
L.~Bai, C.-X. Wang, J.~Huang, Q.~Xu, Y.~Yang, G.~Goussetis, J.~Sun, and
  W.~Zhang, ``Predicting wireless mm{W}ave massive {MIMO} channel
  characteristics using machine learning algorithms,'' \emph{Wirel. Commun.
  Mob. Comput.}, vol. 2018, Aug. 2018.

\bibitem{wen2018deep}
C.-K. Wen, W.-T. Shih, and S.~Jin, ``Deep learning for massive {MIMO} {CSI}
  feedback,'' \emph{IEEE Wireless Commun. Lett.}, vol.~7, no.~5, pp. 748--751,
  Oct. 2018.

\bibitem{huang2018big}
J.~Huang, C.-X. Wang, L.~Bai, J.~Sun, Y.~Yang, J.~Li, O.~Tirkkonen, and
  M.~Zhou, ``A big data enabled channel model for 5{G} wireless communication
  systems,'' \emph{IEEE Trans. Big Data}, vol.~6, no.~5, Mar. 2020.

\bibitem{Chen2019}
Z. Zhang, H. Chen, M. Hua, C. Li, Y. Huang and L Yang, ``Double coded caching in ultra dense networks: Caching and multicast scheduling via deep reinforcement learning'', \emph{IEEE
  Trans. Commun. (Early Access)}, Nov. 2019.


\bibitem{zhang2019standard}
L.~Zhang, W.~Zhang, Y.~Li, J.~Sun, and C.-X. Wang, ``Standard condition number
  of hessian matrix for neural networks,'' in \emph{Proc. IEEE Int. Conf.
  Commun. (ICC)}, {Shanghai, China,} May 2019, pp. 1--6.

\bibitem{pennington2017geometry}
J.~Pennington and Y.~Bahri, ``Geometry of neural network loss surfaces via
  random matrix theory,'' in \emph{Proc. Int. Conf. Mach. Learn. (ICML)}, {Sydney, Australia,} Aug.
  2017, pp. 2798--2806.

\bibitem{louart2018random}
C.~Louart, Z.~Liao, R.~Couillet \emph{et~al.}, ``A random matrix approach to
  neural networks,'' \emph{{Ann. Appl. Probab.}}, vol.~28, no.~2,
  pp. 1190--1248, {Apr.} 2018.

\bibitem{nguyen2018loss}
Q.~Nguyen and M.~Hein, ``The loss surface and expressivity of deep
  convolutional neural networks,'' \emph{arXiv preprint arXiv:1710.10928}, {Oct. 2017}.

\bibitem{Wang2019}
H. Wang, Y. Wu, G. Min, J. Xu, and P. Tang, ``Data-driven dynamic resource scheduling for network slicing: A deep reinforcement learning approach,'' \emph{Information Sciences}, vol. 498, pp. 106--116, Sept. 2019.

\bibitem{hu2018deep}
X.~Hu, S.~Liu, R.~Chen, W.~Wang, and C.~Wang, ``A deep reinforcement
  learning-based framework for dynamic resource allocation in multibeam
  satellite systems,'' \emph{IEEE Commun. Lett.}, vol.~22, no.~8, pp.
  1612--1615, Aug. 2018.

\bibitem{liu2018deepnap}
J.~Liu, B.~Krishnamachari, S.~Zhou, and Z.~Niu, ``Deep{N}ap: {D}ata-driven base
  station sleeping operations through deep reinforcement learning,'' \emph{IEEE
  Internet Things J.}, vol.~5, no.~6, pp. 4273--4282, Dec. 2018.

\bibitem{zhu2018deep}
H.~Zhu, Y.~Cao, W.~Wang, T.~Jiang, and S.~Jin, ``Deep reinforcement learning
  for mobile edge caching: Review, new features, and open issues,'' \emph{IEEE
  Netw.}, vol.~32, no.~6, pp. 50--57, Nov./Dec. 2018.

\bibitem{paper7}
S.~Wang, H.~Liu, P.~H. Gomes, and B.~Krishnamachari,
  ``\BIBforeignlanguage{English}{Deep reinforcement learning for dynamic
  multichannel access in wireless networks},''
  \emph{\BIBforeignlanguage{English}{IEEE Trans. Cogn. Commun. Netw.}}, vol.~4,
  no.~2, pp. 257--265, Jun. 2018.

\bibitem{paper8}
C.~Zhong, Z.~Lu, M.~C. Gursoy, and S.~Velipasalar, ``Actor-{C}ritic deep
  reinforcement learning for dynamic multichannel access,'' in \emph{Proc. IEEE
  Glob. Conf. Signal Inf. Process. (GlobalSIP)}, {Anaheim, USA,} Nov. 2018, pp. 599--603.

\bibitem{paper8trans}
C.~{Zhong}, Z.~{Lu}, M.~C. {Gursoy}, and S.~{Velipasalar}, ``A deep
  {A}ctor-{C}ritic reinforcement learning framework for dynamic multichannel
  access,'' \emph{arXiv preprint arXiv:1908.08401}, {Aug.} 2019.

\bibitem{paper9}
H.~Q. Nguyen, B.~T. Nguyen, T.~Q. Dong, D.~T. Ngo, and T.~A. Nguyen, ``Deep
  {Q}-learning with multiband sensing for dynamic spectrum access,'' in
  \emph{Proc. IEEE Int. Symp. Dyn. Spectr. Access Netw. (DySPAN)}, {Seoul, South Korea,} Oct. 2018,
  pp. 1--5.

\bibitem{Chang2019Distributive}
H.~{Chang}, H.~{Song}, Y.~{Yi}, J.~{Zhang}, H.~{He}, and L.~{Liu},
  ``Distributive dynamic spectrum access through deep reinforcement learning: A
  reservoir computing-based approach,'' \emph{IEEE Internet Things J.}, vol.~6,
  no.~2, pp. 1938--1948, Apr. 2019.

\bibitem{paper10}
O.~Naparstek and K.~Cohen, ``\BIBforeignlanguage{English}{Deep multi-user
  reinforcement learning for distributed dynamic spectrum access},''
  \emph{\BIBforeignlanguage{English}{IEEE Trans. Wireless Commun.}}, vol.~18,
  no.~1, pp. 310--323, Jan. 2019.

\bibitem{paper11}
Y.~Yu, T.~Wang, and S.~C. Liew,
  ``\BIBforeignlanguage{English}{Deep-reinforcement learning multiple access
  for heterogeneous wireless networks},''
  \emph{\BIBforeignlanguage{English}{IEEE J. Sel. Areas Commun.}}, vol.~37,
  no.~6, pp. 1277--1290, Jun. 2019.

\bibitem{paper12}
X.~Liu, Y.~Xu, L.~Jia, Q.~Wu, and A.~Anpalagan,
  ``\BIBforeignlanguage{English}{Anti-jamming communications using spectrum
  waterfall: A deep reinforcement learning approach},''
  \emph{\BIBforeignlanguage{English}{IEEE Commun. Lett.}}, vol.~22, no.~5, pp.
  998--1001, May 2018.

\bibitem{paper13}
P.~Yang, L.~Li, J.~Yin, H.~Zhang, W.~Liang, W.~Chen, and Z.~Han, ``Dynamic
  spectrum access in cognitive radio networks using deep reinforcement learning
  and evolutionary game,'' in \emph{Proc. IEEE/CIC Int. Conf. Commun. in China
  (ICCC)}, {Beijing, China,} Aug. 2018, pp. 405--409.

\bibitem{paper14}
V.~Mnih, K.~Kavukcuoglu, D.~Silver, A.~A. Rusu, J.~Veness, M.~G. Bellemare,
  A.~Graves, M.~Riedmiller, A.~K. Fidjeland, G.~Ostrovski, S.~Petersen,
  C.~Beattie, A.~Sadik, I.~Antonoglou, H.~King, D.~Kumaran, D.~Wierstra,
  S.~Legg, and D.~Hassabis, ``\BIBforeignlanguage{English}{Human-level control
  through deep reinforcement learning},''
  \emph{\BIBforeignlanguage{English}{Nature}}, vol. 518, no. 7540, pp.
  529--533, {Feb.} 2015.

\bibitem{fang2017learning}
M.~Fang, Y.~Li, and T.~Cohn, ``Learning how to active learn: A deep
  reinforcement learning approach,'' \emph{arXiv preprint arXiv:1708.02383}, {Aug.}
  2017.

\bibitem{paper15}
D.~Kingma and J.~Ba, ``Adam: A method for stochastic optimization,''
  \emph{{arXiv preprint arXiv:1412.6980}}, {Dec.} 2014.

\end{thebibliography}

\begin{IEEEbiography}[{\includegraphics[width=1in,height=1.25in,clip,keepaspectratio]{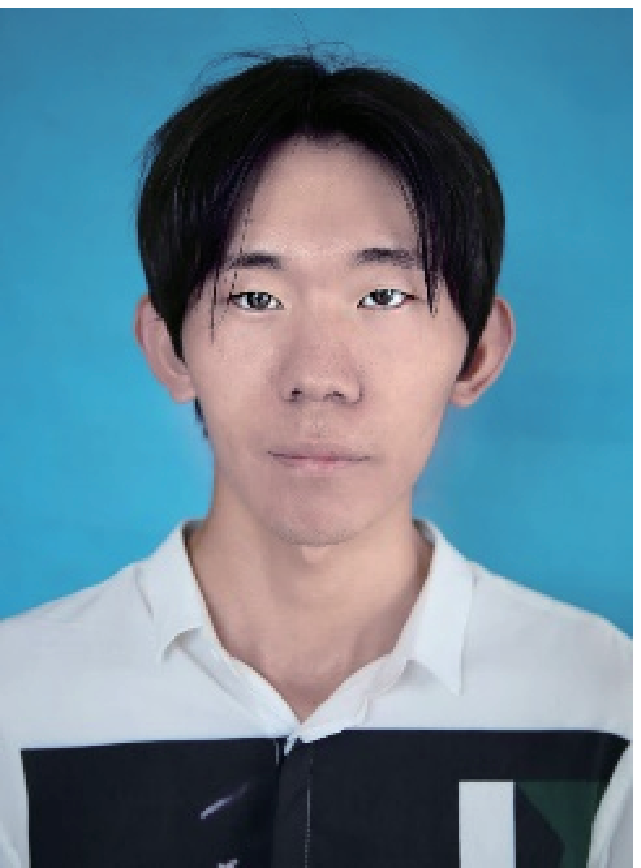}}]
{Yunzeng~Li} received the BSc degree in Communication Engineering from Shandong University, China, in 2018. He is currently pursuing the MEng degree in Electronics and Communication Engineering in Shandong University, China. His research interests include deep reinforcement learning and B5G dynamic spectrum allocation.
\end{IEEEbiography}

\begin{IEEEbiography}[{\includegraphics[width=1in,height=1.25in,clip,keepaspectratio]{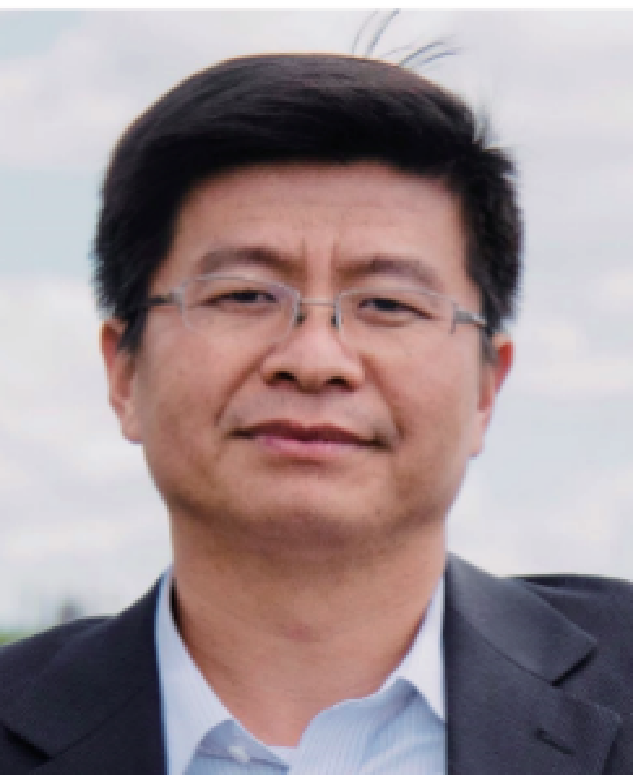}}]
{Wensheng~Zhang} (S'08-M'11) received the M.E. degree and Ph.D. degree both in electrical engineering from Shandong University, China, in 2005 and Keio University, Japan, in 2011, respectively. In 2011, he joined the School of Information Science and Engineering at Shandong University, where he is currently an associate professor. He was a visiting scholar at University of Oulu, Finland, 2010 and University of Arkansas, USA, 2019. His research interests lie in tensor computing, random matrix theory, and intelligent B5G wireless communications.
\end{IEEEbiography}

\begin{IEEEbiography}[{\includegraphics[width=1in,height=1.25in,clip,keepaspectratio]{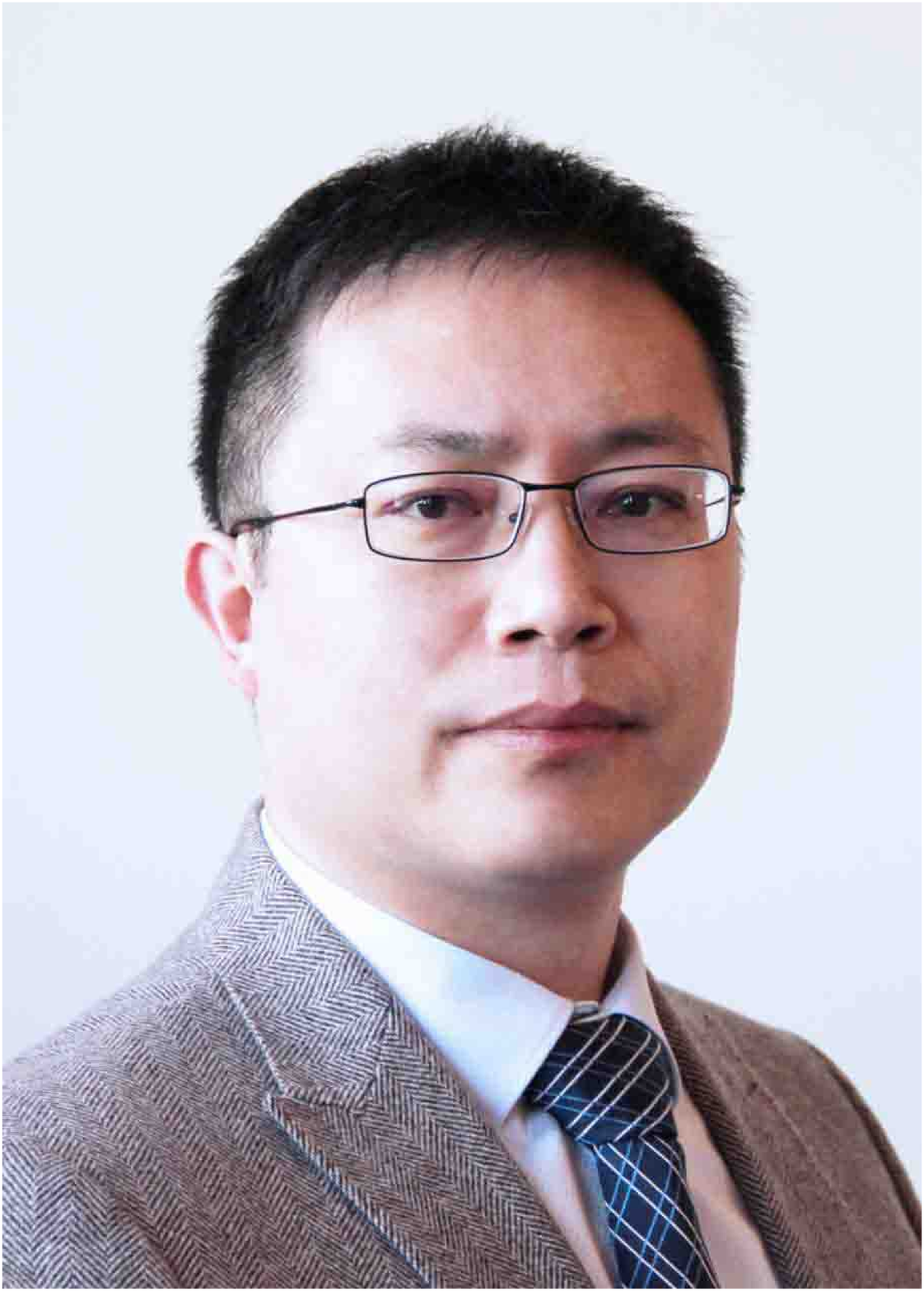}}]
{Cheng-Xiang~Wang} (S’01-M’05-SM’08-F’17) received the BSc and MEng degrees in Communication and Information Systems from Shandong University, China, in 1997 and 2000, respectively, and the PhD degree in Wireless Communications from Aalborg University, Denmark, in 2004.

He was a Research Assistant with the Hamburg University of Technology, Hamburg, Germany, from 2000 to 2001, a Visiting Researcher with Siemens AG Mobile Phones, Munich, Germany, in 2004, and a Research Fellow with the University of Agder, Grimstad, Norway, from 2001 to 2005. He has been with Heriot-Watt University, Edinburgh, U.K., since 2005, where he was promoted to a Professor in 2011. In 2018, he joined Southeast University, China, as a Professor. He is also a part-time professor with the Purple Mountain Laboratories, Nanjing, China. He has authored three books, one book chapter, and more than 370 papers in refereed journals and conference proceedings, including 23 Highly Cited Papers. He has also delivered 18 Invited Keynote Speeches/Talks and 7 Tutorials in international conferences. His current research interests include wireless channel measurements and modeling, B5G wireless communication networks, and applying artificial intelligence to wireless communication networks.

Prof. Wang is a fellow of the IET, an IEEE Communications Society Distinguished Lecturer in 2019 and 2020, and a Highly-Cited Researcher recognized by Clarivate Analytics, in 2017-2019. He is currently an Executive Editorial Committee member for the IEEE TRANSACTIONS ON WIRELESS COMMUNICATIONS. He has served as an Editor for nine international journals, including the IEEE TRANSACTIONS ON WIRELESS COMMUNICATIONS from 2007 to 2009, the IEEE TRANSACTIONS ON VEHICULAR TECHNOLOGY from 2011 to 2017, and the IEEE TRANSACTIONS ON COMMUNICATIONS from 2015 to 2017. He was a Guest Editor for the IEEE JOURNAL ON SELECTED AREAS IN COMMUNICATIONS, Special Issue on Vehicular Communications and Networks (Lead Guest Editor), Special Issue on Spectrum and Energy Efficient Design of Wireless Communication Networks, and Special Issue on Airborne Communication Networks. He was also a Guest Editor for the IEEE TRANSACTIONS ON BIG DATA, Special Issue on Wireless Big Data, and is a Guest Editor for the IEEE TRANSACTIONS ON COGNITIVE COMMUNICATIONS AND NETWORKING, Special Issue on Intelligent Resource Management for 5G and Beyond. He has served as a TPC Member, TPC Chair, and General Chair for over 80 international conferences. He received ten Best Paper Awards from IEEE GLOBECOM 2010, IEEE ICCT 2011, ITST 2012, IEEE VTC 2013-Spring, IWCMC 2015, IWCMC 2016, IEEE/CIC ICCC 2016, WPMC 2016, and WOCC 2019.
\end{IEEEbiography}

\begin{IEEEbiography}[{\includegraphics[width=1in,height=1.25in,clip,keepaspectratio]{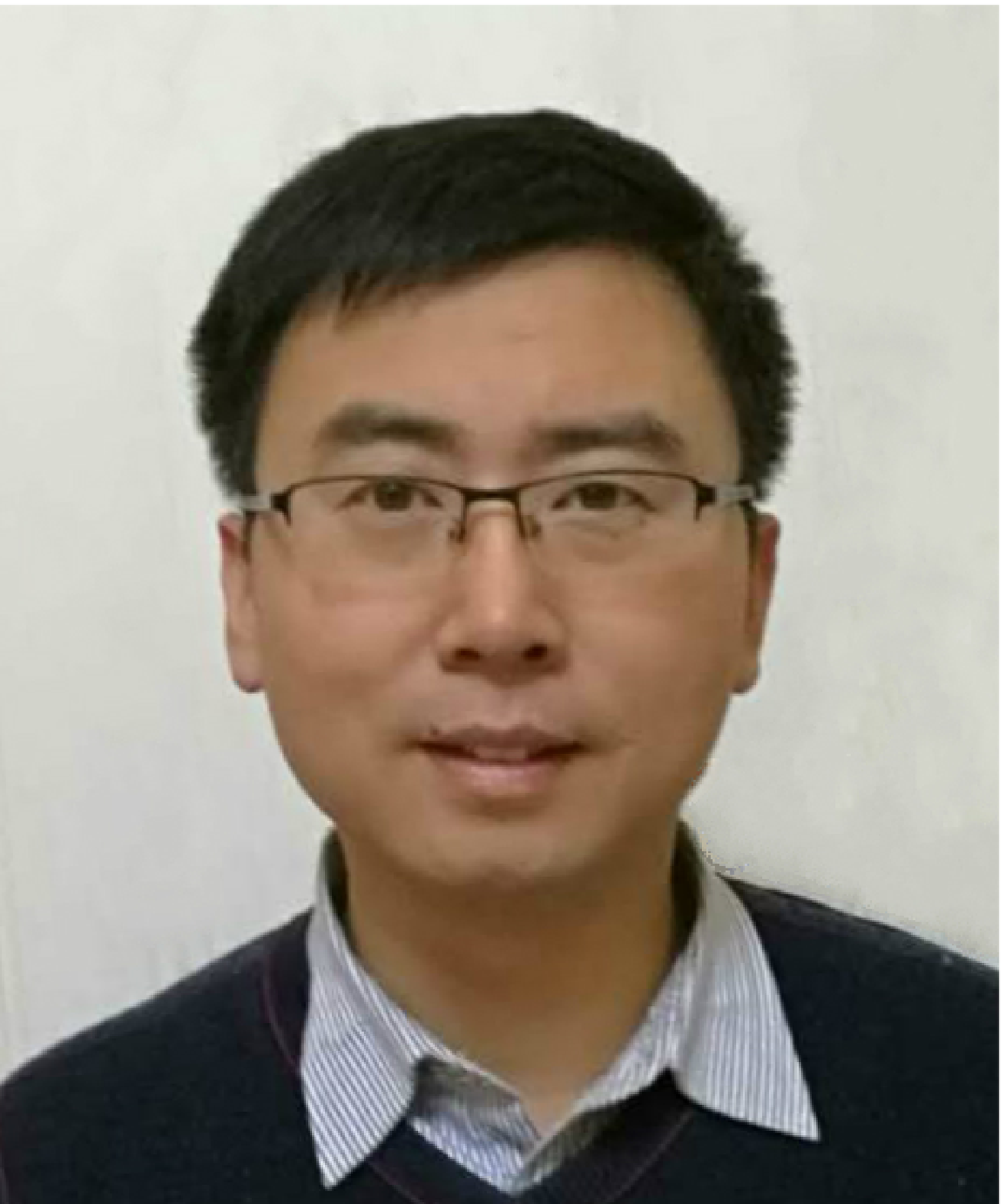}}]
{Jian~Sun} (M'08) received the B.Sc. degree in applied electronic technology, the M.Eng. degree in measuring and testing technologies and instruments, and the Ph.D. degree in communication and information systems, all from Zhejiang University, Hangzhou, China, in 1996, 1999, and 2005, respectively.

From 2005 to 2018, he was a Lecturer with the School of Information Science and Engineering, Shandong University, China.  Since 2018, he has been an Associate Professor. In 2008, he was a Visiting Scholar with University of California San Diego (UCSD). In 2011, he was a Visiting Scholar with Heriot-Watt University, U.K., supported by U.K.–China Science Bridges: R\&D on B4G Wireless Mobile Communications project. His current research interests include signal processing for wireless communications, channel sounding and modeling, propagation measurement and parameter extraction, maritime communication, visible light communication, software defined radio, MIMO, multicarrier, wireless systems design and implementation.
\end{IEEEbiography}

\begin{IEEEbiography}[{\includegraphics[width=1in,height=1.25in,clip,keepaspectratio]{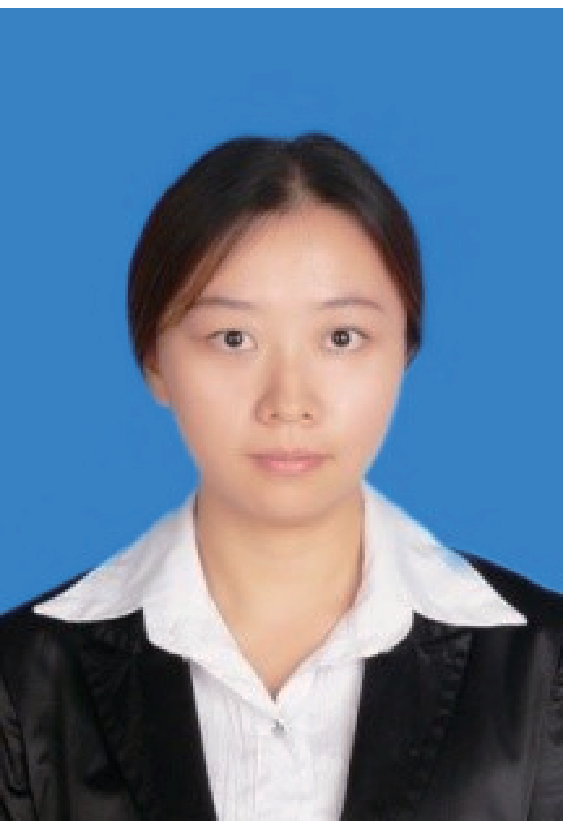}}]
{Yu~Liu} received the Ph.D. degree in communication and information systems from Shandong University, Jinan, China, in 2017. From 2015 to 2017, she was a visiting scholar with the School of Engineering and Physical Sciences, Heriot–Watt University, Edinburgh, U.K.. From 2017 to 2019, she was a Postdoctoral Research Associate with the School of Information Science and Engineering, Shandong University, Jinan, China. Since 2019, she has been an Associate Professor with the School of Microelectronics, Shandong University, Jinan, China. Her main research interests include nonstationary wireless MIMO channel modeling, high speed train wireless propagation characterization and modeling, and channel modeling for special scenarios.
\end{IEEEbiography}

\end{document}